\documentclass[twocolumn]{aastex61}
\pdfoutput=1 
\usepackage{amsmath,amstext}
\usepackage[T1]{fontenc}
\usepackage{apjfonts} 
\usepackage[figure,figure*]{hypcap}
\usepackage{threeparttable}
\usepackage{color}
\usepackage{longtable}

\shorttitle{The relation between outburst rate and orbital period in low-mass X-ray binary transients}
\shortauthors{Jie Lin et al.}

\begin{document}

\title{The relation between outburst rate and orbital period in low-mass X-ray binary transients}



\author{Jie Lin}
\affiliation{Shanghai Astronomical Observatory and Key Laboratory for Research Galaxies and Cosmology, Chinese Academy of Sciences, 80 Nandan Road, Shanghai 200030, China: wenfei@shao.ac.cn}
\affiliation{University of Chinese Academy of Sciences, 19 A Yuquan Road, Beijing 100049, China:  linjie@shao.ac.cn}
\author{Zhen Yan}
\affiliation{Shanghai Astronomical Observatory and Key Laboratory for Research Galaxies and Cosmology, Chinese Academy of Sciences, 80 Nandan Road, Shanghai 200030, China: wenfei@shao.ac.cn}
\author{Zhanwen Han}
\affiliation{Yunnan Observatory, Academia Sinica, Kunming 650011, China}
\author{Wenfei Yu}
\affiliation{Shanghai Astronomical Observatory and Key Laboratory for Research Galaxies and Cosmology, Chinese Academy of Sciences, 80 Nandan Road, Shanghai 200030, China: wenfei@shao.ac.cn}

\begin{abstract}
We have investigated the outburst properties of low-mass X-ray binary transients (LMXBTs) based on a comprehensive study of the outbursts observed in the past few decades.  The outburst rates were estimated based on the X-ray monitoring data from \emph{Swift}/BAT, RXTE/ASM and MAXI, and previous reports in the literature. We found that almost all LMXBTs with the orbital period below $\sim$12 hr showed only one outburst in these observations. There are systematic difference in the outburst rate  between long-period ($P_{\rm orb} \gtrsim$ 12 hr) and short-period ($P_{\rm orb} \lesssim$ 12 hr) systems. We infer that mass transfer rate is responsible for the systematic difference, since the disk instability model (DIM) suggested that the mass transfer rate is a key factor affecting the quiescence time. The difference in outburst rate between long-period and short-period LMXBTs is probably due to the different mass transfer mechanism at different evolutionary stages of the donors. Based on the evolutionary tracks of single stars, we derived the critical orbital period for X-ray binaries that harbor a subgiant donor in various metallicity. The critical orbital period ($P_{\rm orb,crit}=$12.4 hr) is consistent with the above orbital period boundary obtained from the statistics of outburst rates. Furthermore, we found a negative correlation between the outburst rate and the orbital period in the samples for which the luminosity class of the donor star is III/IV. The best-fitting power-law index for the black hole subsamples is roughly consistent with the theoretical prediction for those systems with a donor star evolved off the main sequence. 

\end{abstract}


\section{Introduction}
Low-mass X-ray binaries (LMXBs) are binary systems in which a donor star transfers mass onto a black hole (BH) or a neutron star (NS) through Roche lobe overflows. The donor star is usually a main-sequence star, a subgiant, or even a white dwarf (WD). Among LMXBs, low-mass X-ray binary transients (LMXBTs) spend most of their time in X-ray quiescence (typically $L\leq 10^{34}~{\rm erg~s^{-1}}$) and occasionally turn into outbursts, during which their X-ray luminosities can increase by many orders of magnitude (see \citealt{Chen+etal+1997}, hereafter CSL97). 

Similar to those of dwarf novae, the outburst behavior of LMXBTs is usually believed to be the result of disk instability (\citealt{Lasota+2001,Osaki+1995}). 
In the disk instability model (DIM), the trigger mechanism is the thermal instability within the accretion disk. 
The disk thermal instability occurs when the surface density exceeds the maximum value on the cold branch or the surface density is lower than the minimum value on the hot branch (\citealt{Lasota+2001,Lasota+2016}).

When and where the instability is triggered are usually related to the following two processes: mass accumulation at the outer edge of the disk, and the viscous diffusion of the matter accumulated in the outer disk. Two corresponding timescales matter.  The mass-accumulation time is the time duration for accumulating mass up to the critical surface density of the cold branch at the outer disk. The viscous drift time is the time for the matter in the disk to drift appreciably in the radial direction and trigger the disk instability in the inner disk. Hence, theoretically, the duration for a transient to stay in quiescence is determined by the shorter one of the two timescales (\citealt{Smak+1984,Osaki+1995,Lasota+2001}).
According to \cite{Osaki+1995}, the mass-accumulation time depends on the mass transfer rate and the outer disk radius, while the viscous drift time is independent of the mass transfer rate from the companion star and is weakly dependent on the disk outer radius. Therefore, the mass transfer rate and outer disk radius are thought to be the key factors that determine whether the matter will tend to accumulate on the outer edge or propagate inward (\citealt{Lasota+2001,Dubus+etal+2001}). The mass transfer rate is tightly related with the orbital period, mass, and the evolutionary stage of the donor (\citealt{Webbink+etal+1983,King+1988}). On the other hand, the outer disk radius is also positively correlated with the orbital period in LMXBs (\citealt{Lasota+2001,Lasota+etal+2008}). Therefore, the quiescence time might be related to the orbital periods, as well as the nature of donors.

From the observational point of view, the quiescence time of LMXBTs is difficult to measure owing to the limited sensitivity of X-ray monitors, and we cannot determine when a transient enters or leaves the quiescence state exactly. Most recent all-sky monitors offer good time coverage, but their sensitivities are still far above the flux level of the quiescence. On the other hand, the outburst rate, i.e., the number of outbursts per year, which is the reciprocal of the average recurrence time, can be counted directly when a flux threshold is taken. Notice that the quiescence time is the time duration from the end of an outburst to the start of the next one, which is shorter than the recurrence time, defined as the time interval from the start of an outburst to the start of the next outburst (\citealt{Lasota+2001}). For most LMXBTs now known, the outburst duty cycle is quite small (\citealt{Yan+Yu+2015},hereafter YY15;\citealt{Tetarenko+etal+2016}, hereafter TSHG16), so the quiescence time and the recurrence time in the current sample are usually comparable. In the following, we present our measurements of the outburst rates for all known LMXBTs in which the orbital period has been determined, aiming at obtaining the potential relation between the recurrence time and the orbital period or the properties of the donor.

\subsection{Case A and Case B mass transfer}

The most common mechanism driving mass transfer in a binary system is the expansion of the donor (see \citealt{Pylyser+Savonije+1988}). The radius of an intermediate-mass star (typically 3 $\rm M_{\odot}$) grows in three different epochs: the phase of core-hydrogen burning (main-sequence stage), the phase before ignition of core helium burning (subgiant branch and red giant branch) and the phase after termination of core helium burning (asymptotic giant branch). The mass transfer in these three phases is often referred to as case A, case B, and case C mass transfer, respectively (see also \citealt{Kolb+book+2010}). Notice that, for low-mass stars, their main-sequence lifetime could be longer than the age of the universe (or Hubble time), and they never become (sub)giants if their initial masses are small enough ($M_{\rm 2,i}\lesssim0.9~{\rm M_{\odot}}$).
The radius of a donor star in case C mass transfer is much larger than those in case B and case A, which implies a much larger orbital separation and orbital period. Such an orbital period exceeds the maximum orbital period of known LMXBTs; therefore, case C mass transfer is beyond the scope of our study.

In case B mass transfer, the radius grows on thermal time $t_{\rm th}$ of the donor star, which is approximately equal to the Kelvin-Helmholtz time.
In case A mass transfer, the radius of the donor grows on its nuclear time $t_{\rm nuc}$, which is typically a factor of 1000 times longer than the thermal time $t_{\rm th}$. 
Therefore, the case B mass transfer rate is much larger than the case A mass transfer rate. Typically, the case B mass transfer rate can reach $10^{-7}~{\rm M_{\odot}/yr}$, while the case A mass transfer rate is about several times $10^{-11}~{\rm M_{\odot}/yr}$ (see \citealt{Kool+etal+1986,Kolb+book+2010}).
Since the size of (sub)giants is usually several times or tens of times that of main-sequence stars, only those LMXBTs with a sufficiently large orbital period can harbor a donor of a (sub)giant. In other words, case B mass transfer can be performed only if the LMXBTs have an orbital period greater than a \emph{critical orbital period}. In this paper, we will discuss the relation between our results for the outburst rate and the critical orbital period.

\label{introduction}

\section{Data}\label{data}
We selected the LMXBT samples with confirmed measurements of the orbital periods based on the catalog of \cite{Knevitt+etal+2014} and \cite{Tetarenko+etal+2016}. We then searched for the X-ray outbursts of these samples mainly from the archived data of the all-sky monitors, including RXTE/ASM, \emph{Swift}/BAT and MAXI, since they are complete for detecting X-ray outbursts of LMXBTs in the past two decades (YY15; TSHG16). We also searched for outbursting events in the literature before the launch of RXTE. Due to limited sensitivity and some poor time coverage of the all-sky X-ray monitors, it is impossible to identify all the outbursts of LMXBTs in the period we investigated.  However, we can find all the outbursts with the X-ray peak flux above a certain threshold from the all-sky monitoring data. Considering the typical 1-day X-ray sensitivity (10--20 mcrab) of RXTE/ASM, \emph{Swift}/BAT and MAXI, we set the threshold to 50 mcrab of our outburst sample, corresponding to about $10^{37}$ erg s$^{-1}$ at 2--10 keV if a source distance of 8 kpc is taken. 

To ensure the completeness of  sample, our samples include all LMXBTs with known orbital periods except the following sources.
\emph{Swift} J1357.2--0933, XTE J0929--314, XTE J1710--281, IGR J17498--2921, XTE J1814--338, 4U 2129+47 (\citealt{Forman+Jokipii+1978,Bozzo+etal+2007}) and SWIFT J1749.4--2807 are excluded from our samples owing to their weak outbursts (below 50 mcrab). AX J1745.6--2901 is also excluded because it is in the Galactic center and may be confused with another X-ray source (e.g. CXOGC J174540.0-290027 and CXOGC J174540.0-290031). All our LMXBT samples are listed in Table~\ref{table1}.

We have examined the orbital periods of our samples. As shown in Table~\ref{table1}, the measurements of the orbital periods are limited to four common measurement methods: the Doppler-shifted absorption (or emission) lines, the periodic modulations in optical and IR band, the periodic dips in the X-ray emission, and the Doppler-delayed pulses. In our samples, MAXI J0556--332 has two candidate orbital periods (16.43 and 9.754 hr). The orbital period of 4U 0042+32 may be not be reliable, since it is measured by periodic X-ray modulation rather than the X-ray dips in the common cases. For example, \cite{Kaluzienski+etal+1980} also found an 8.2-hour periodic X-ray modulation in Cen X-4, but its orbital period is believed to be 15.1 hr which was measured by the Doppler-shifted emission lines (\citealt{Cowley+etal+1988}) and the periodic modulation in V band (\citealt{Chevalier+1989}).

In order to investigate the evolutionary stage of the donors in our samples, we collected the luminosity classes of the donors from the literature.  Typically, the luminosity class notation ``V'' indicates that the star is at the main-sequence stage, and the notation ``III'' or ``IV'' indicates that the star is at (sub)giant stage. The luminosity class of the donor could be jointly determined by the absolute magnitude and the spectra (or by the width of spectral lines). Notice that the optical and infrared emission from X-ray binaries might contain the contribution from the accretion disk or jet even during the quiescence, which means that luminosity class may be not a very reliable indicator to classify the donors. We listed all the known spectral classifications of the donors in Table~\ref{table1}.

\subsection{The number of outbursts}

Firstly, we describe here how we selected the outburst samples (during or after the RXTE era) from the X-ray monitoring data in detail. We retrieved the orbital or dwell-by-dwell light curves from the public archives of RXTE/ASM, \emph{Swift}/BAT and MAXI (see, e.g. \citealt{Levine+etal+1996,Matsuoka+etal+2009,Krimm+etal+2013}).  Among these, the light curves from RXTE/ASM in 2--12 keV correspond to the period from MJD 50088 to $\sim$MJD 55924, the light curves from \emph{Swift}/BAT in 15--50 keV correspond to the period from MJD 53414 to MJD 57900 and the light curves from MAXI in 2--4 keV and 4--10 keV correspond to the period from MJD 55058 to MJD 57900. We excluded the MAXI data in the 10--20 keV energy band due to its relatively poor signal-to-noise ratio as compared to \emph{Swift}/BAT.  All light curves were rebinned into daily averaged data, and all the X-ray intensity is converted into intensity in crab unit for the purpose of comparison among data from different instruments. We used $1~{\rm Crab} = 75.6~{\rm count~s^{-1}}$ for RXTE/ASM, $1~{\rm Crab} = 0.221~{\rm count~s^{-1}~cm^{-2}}$ for \emph{Swift}/BAT, $1~{\rm Crab} = 1.67~{\rm count~s^{-1}~cm^{-2}}$ and $1.15~{\rm count~s^{-1}~cm^{-2}}$ for MAXI at 2--4 keV and 4--10 keV energy bands, respectively.

We then searched for any outbursts with the X-ray peak flux above 50 mcrab from the long-term light curves of different instruments. In order to avoid from selecting fake outbursts (e.g. the type I X-ray burst in NS XRBs), we applied the criterion by requiring that there should be at least three successive data points with signal-to-noise ratio better than 3$\sigma$, i.e., the X-ray outbursts from our LMXBT samples all lasted more than 2 days. 

\renewcommand\tabcolsep{1pt}
\begin{longtable*}{lccccccccc}
\caption[]{Low-mass X-ray Binary Transients Information}  \\
\hline \hline \\[-4ex]
   \multicolumn{1}{l}{Source Name} &
   \multicolumn{1}{c}{Orbital Period$^a$} &
     \multicolumn{1}{c}{Method$^a$} &
     \multicolumn{1}{c}{Spectral Type$^b$} &
     \multicolumn{1}{c}{Discovery} &
    \multicolumn{4}{c}{Number of Outbursts$^c$}&
   \multicolumn{1}{c}{Rate$^f$}\\
   \cline{6-9}
&(hr)	&	&& (MJD)	&	Era I$^d$	& Era II$^e$ & Era III&Total&(yr$^{-1}$)	\\[-3ex]
\label{table1}
\endfirsthead

\multicolumn{9}{c}{{\tablename} \thetable{} -- Continued} \\[0.5ex]
\hline \hline \\[-4ex]
   \multicolumn{1}{l}{Source Name} &
   \multicolumn{1}{c}{Orbital Period$^a$} &
     \multicolumn{1}{c}{Method$^a$} &
   \multicolumn{1}{c}{Spectral Type$^b$} &
     \multicolumn{1}{c}{Discovery} &
    \multicolumn{4}{c}{Number of Outbursts$^c$}&
   \multicolumn{1}{c}{Rate$^f$}\\
   \cline{6-9}
&(hr)	&	&& (MJD)	&	Era I$^d$	& Era II$^e$ & Era III&Total&(yr$^{-1}$)	\\[0.5ex]
\hline\\[-3ex]
\endhead

  \\[-2ex] \hline  \\[-1ex]
\endfoot
  \\[-2ex] \hline \\[-1.5ex]  
    \multicolumn{10}{p{\columnwidth}}{ {\bf Notes.} 
    $^a$ The methods used for the orbital period measurements: D (periodic dips), M (periodic modulations), A (Doppler-shifted absorption or emission lines), and P (Doppler-delayed pulses).
    }\\[-0.5ex]
    \multicolumn{10}{p{\columnwidth}}{$^b$ The spectral types here depend on the spectroscopy and photometry from optical or infrared observations.
    }\\[-0.5ex]
    \multicolumn{10}{p{\columnwidth}}{$^c$ We list the number of outbursts in era I (before 1996), era II (from early to late 2011, the RXTE era), and era III (after 2011). For the outbursts that span multiple periods, we attribute them to the period in which they began to outburst. For the BH LMXBTs, we also list the numbers based on the outburst list of TSHG16 and use ``/'' to divide our and their results. 
    }\\[-0.5ex]
    \multicolumn{10}{p{\columnwidth}}{$^d$ The numbers in the parentheses represent the numbers of the outbursts (above 50 mcrab) recorded by CSL97 for the outbursts before the RXTE era.
    }\\[-0.5ex]
    \multicolumn{10}{p{\columnwidth}}{$^e$ The numbers in the parentheses show the numbers of the outbursts that are counted by YY15 for the outbursts in the RXTE era.
    }\\[-0.5ex]
    \multicolumn{10}{p{\columnwidth}}{
    $^f$ We set the upper limit on the outburst rate for those LMXBTs in which only one outburst has been detected above the threshold during the entire history of X-ray observations.
    }\\[-0.5ex]
    \multicolumn{10}{p{\columnwidth}}{
    $^*$ An extra outburst of 1A 1744--361 is based on the report of the MAXI team when it is not in the monitoring catalog of MAXI (\citealt{atel+5301}).
	$^{**}$ The total numbers of outbursts for 1A 1744--361 and GRS 1747--312 are counted since the RXTE era (early 1996).
    $^{***}$The orbital period of 4U 0042+32 was measured by periodic X-ray modulation rather than the X-ray dips in the common cases.
    }\\
 \multicolumn{10}{p{\columnwidth}}
 {{\bf References.} 
[1] \cite{Kuulkers+etal+2013}; [2] \cite{atel2873+2010}; [3] \cite{Zurita+etal+2008}; [4] \cite{Neustroev+etal+2014}; [5] \cite{atel546+2005}; [6] \cite{McClintock+etal+2003}; [7] \cite{Gelino+etal+2006}; [8] \cite{iauc7389+2000}; [9] \cite{Filippenko+etal+1995}; [10] \cite{Gelino+Harrison+2003}; [11] \cite{iauc5580+1992}; [12] \cite{iauc7303+1999}; [13] \cite{Corral+etal+2011}; [14] \cite{iauc7274+1999}; [15] \cite{Shahbaz+etal+1996}; [16] \cite{Filippenko+etal+1999}; [17] \cite{iauc5864+1993}; [18] \cite{Orosz+etal+2004}; [19] \cite{iauc7707+2001}; [20] \cite{McClintock+etal+1983}; [21] \cite{Shahbaz+etal+1999b}; [22] \cite{Gelino+etal+2001}; [23] \cite{iauc2814+1975}; [24] \cite{Casares+etal+1995}; [25] \cite{iauc4587+1988}; [26] \cite{iauc4583+1988}; [27] \cite{Shidatsu+etal+2013}; [28] \cite{atel4024+2012}; [29] \cite{Orosz+etal+1996}; [30] \cite{Chaty+etal+2002}; [31] \cite{iauc5161+1991}; [32] \cite{Filippenko+etal+1997}; [33] \cite{Harlaftis+etal+1997}; [34] \cite{iauc3104+1977}; [35] \cite{iauc3110+1977}; [36] \cite{Masetti+etal+1996}; [37] \cite{Valle+etal+1994}; [38] \cite{iauc5874+1993}; [39] \cite{iauc6083+1994}; [40] \cite{iauc6104+1994}; [41] \cite{Orosz+etal+1998}; [42] \cite{Chevalier+1989}; [43] \cite{Matilsky+etal+1972}; [44] \cite{Kitamoto+etal+1984}; [45] \cite{iauc5504+1992}; [46] \cite{Orosz+etal+2002}; [47] \cite{Orosz+etal+2011}; [48] \cite{iauc7008+1998}; [49] \cite{Hynes+Jones+2008}; [50] \cite{Markert+etal+1973}; [51] \cite{Kong+etal+2002}; [52] \cite{Casares+etal+2009}; [53] \cite{Casares+etal+2004}; [54] \cite{iauc4342+1987}; [55] \cite{Orosz+Bailyn+1997}; [56] \cite{Greene+etal+2001}; [57] \cite{Shahbaz+etal+1999a}; [58] \cite{iauc6046+1994}; [59] \cite{Casares+etal+1992}; [60] \cite{Casares+Charles+1994}; [61] \cite{iauc4782+1989}; [62] \cite{Neil+etal+2007}; [63] \cite{Harlaftis+Greiner+2004}; [64] \cite{iauc5590+1992}; [65] \cite{Chou+etal+2008}; [66] \cite{atel122+2003}; [67] \cite{Markwardt+etal+2002}; [68] \cite{iauc7867+2002}; [69] \cite{Krimm+etal+2007}; [70] \cite{atel1105+2007}; [71] \cite{Kaaret+etal+2005}; [72] \cite{atel516+2005}; [73] \cite{Bhattacharyya+etal+2006}; [74] \cite{iauc2925+1976}; [75] \cite{atel268+2004}; [76] \cite{atel+5301}; [77] \cite{Chakrabarty+Morgan+1998}; [78] \cite{Zand+etal+1998}; [79] \cite{Homer+etal+1998}; [80] \cite{Meshcheryakov+etal+2010}; [81] \cite{iauc4653+1988}; [82] \cite{Galloway+etal+2005}; [83] \cite{atel352+2004}; [84] \cite{Papitto+etal+2010}; [85] \cite{atel2196+2009}; [86] \cite{Parmar+etal+1986}; [87] \cite{Hynes+Jones+2009}; [88] \cite{Ratti+etal+2012}; [89] \cite{iauc4039+1985}; [90] \cite{Tomsick+etal+1999}; [91] \cite{Casares+etal+2002}; [92] \cite{Tomsick+etal+2002}; [93] \cite{iauc6955+1998}; [94] \cite{Cominsky+Wood+1989}; [95] \cite{Wachter+Smale+1998}; [96] \cite{iauc2994+1976}; [97] \cite{Zand+etal+2003}; [98] \cite{Pavlinsky+etal+1994}; [99] \cite{Wachter+etal+2002}; [100] \cite{Lochner+Roussel+1994}; [101] \cite{iauc2859+1975}; [102] \cite{iauc3349+1979}; [103] \cite{Mitsuda+etal+1989}; [104] \cite{Zhang+etal+1996}; [105] \cite{Cowley+etal+1988}; [106] \cite{Chevalier+etal+1989}; [107] \cite{Paradijs+etal+1980}; [108] \cite{Baglio+etal+2014}; [109] \cite{Conner+etal+1969}; [110] \cite{iauc3360+1979}; [111] \cite{Matsuoka+etal+1980}; [112] \cite{Cornelisse+etal+2012}; [113] \cite{atel3102+2011}; [114] \cite{Chevalier+Ilovaisky+1998}; [115] \cite{Welsh+etal+2000}; [116] \cite{Chevalier+etal+1999}; [117] \cite{Kitamoto+etal+1993}; [118] \cite{Harmon+etal+1996}; [119] \cite{Watson+Ricketts+1978}; [120] \cite{Charles+etal+1978}; [121] \cite{Laros+Wheaton+1980}; [122] \cite{Kouveliotou+etal+1996}; [123] \cite{Gosling+etal+2007}; [124] \cite{iauc6272+1995}.
        }\\                
\endlastfoot
\hline	
\multicolumn{10}{c}{BH LMXBTs}	\\
\hline	
MAXI J1659--152 & 2.41$^{[1]}$ & D & - & 55464$^{[2]}$ & - & 1/1~(1) & 0/0 & 1/1 & $\leq$0.047 \\ 
\emph{Swift} J1753.5--0127 & 3.2$^{[3,4]}$ & A+M & - & 53551$^{[5]}$ & - & 1/1~(1) & 0/0 & 1/1 & $\leq$0.047 \\ 
XTE J1118+480 & 4.1$^{[6,7]}$ & A+M & K5-7V$^{[6,7]}$ & 51608$^{[8]}$ & - & 1/2 & 0/0 & 1/2 & $\leq$0.047 \\ 
GRO J0422+32 & 5.1$^{[9]}$ & A+M & M1(+/-1)V$^{[9,10]}$ & 48839$^{[11]}$ & 1/3~(1)$^{[11]}$ & 0/0 & 0/0 & 1/3 & $\leq$0.040 \\ 
XTE J1859+226 & 6.6$^{[12,13]}$ & A+M & K5-7V$^{[13]}$ & 51460$^{[14]}$ & - & 1/1~(1) & 0/0 & 1/1 & $\leq$0.047 \\ 
GRS 1009--45 & 6.8$^{[15,16]}$ & A+M & K6-M0V$^{[16]}$ & 49242$^{[17]}$ & 1/1~(1)$^{[17]}$ & 0/0 & 0/0 & 1/1 & $\leq$0.042 \\ 
XTE J1650--500 & 7.7$^{[18]}$ & M & K4V$^{[18]}$ & 52157$^{[19]}$ & - & 1/1~(1) & 0/0 & 1/1 & $\leq$0.047 \\ 
1A 0620--00 & 7.8$^{[20]}$ & M & K3-4V$^{[21,22]}$ & 42627$^{[23]}$ & 1/1~(1)$^{[23]}$ & 0/0 & 0/0 & 1/1 & $\leq$0.024 \\ 
GS 2000+251 & 8.3$^{[24]}$ & A & K5(+/-2)V$^{[24]}$ & 47277$^{[25]}$ & 1/1~(1)$^{[26,25]}$ & 0/0 & 0/0 & 1/1 & $\leq$0.034 \\ 
MAXI J1305--704 & 9.74$^{[27]}$ & D & - & 56026$^{[28]}$ & - & 0/0 & 1/1 & 1/1 & $\leq$0.047 \\ 
GS 1124--684 & 10.4$^{[29]}$ & A+M & K0-5V$^{[29,30]}$ & 48264$^{[31]}$ & 1/1~(1)$^{[31]}$ & 0/0 & 0/0 & 1/1 & $\leq$0.038 \\ 
H 1705--250 & 12.5$^{[32]}$ & A & K5-7V$^{[32,33]}$ & 43386$^{[34]}$ & 1/1~(1)$^{[34,35]}$ & 0/0 & 0/0 & 1/1 & $\leq$0.025 \\ 
GRS 1716--249 & 14.7$^{[36]}$ & M & M0-5V$^{[37,30]}$ & 49255$^{[38]}$ & 2/2~(1)$^{[38,39,40]}$ & 0/0 & 1/0 & 3/2 & 0.127 \\ 
4U 1543--47 & 26.8$^{[41]}$ & A+M & A2V$^{[42,41]}$ & 41180$^{[43]}$ & 3/3~(3)$^{[43,44,45]}$ & 1/1~(1) & 0/0 & 4/4 & 0.087 \\ 
XTE J1550--564 & 37.0$^{[46,47]}$ & A & K3III$^{[46,47]}$ & 51063$^{[48]}$ & - & 5/5~(2) & 0/0 & 5/5 & 0.234 \\ 
GX 339--4 & 42.1$^{[49]}$ & A & F8-G2III$^{[30]}$ & $\sim$41360$^{[50,51]}$ & 11/10$^{[50,51]}$ & 7/8~(5) & 2/2 & 20/20 & 0.441 \\ 
GS 1354--64 & 61.1$^{[52,53]}$ & A & G0-5III$^{[53]}$ & 46851$^{[54]}$ & 3/3~(3)$^{[54]}$ & 1/1 & 1/1 & 5/5 & 0.165 \\ 
GRO J1655--40 & 62.9$^{[55,56]}$ & A+M & F6III$^{[55,57]}$ & 49560$^{[58]}$ & 2/1~(2)$^{[58]}$ & 2/2~(2) & 0/0 & 4/3 & 0.175 \\ 
GS 2023+338 & 155.3$^{[59]}$ & A & K0IV$^{[60]}$ & 47668$^{[61]}$ & 1/1~(1)$^{[61]}$ & 0/0 & 1/1 & 2/2 & 0.071 \\ 
GRS 1915+105 & 739.0$^{[62]}$ & M & K1-5III?$^{[63]}$ & 48849$^{[64]}$ & 1/1~(1)$^{[64]}$ & 0/0 & 0/0 & 1/1 & $\leq$0.040 \\ 
\hline	
\multicolumn{10}{c}{NS LMXBTs}	\\
\hline	
XTE J1807--294 & 0.67$^{[65]}$ & P & - & 52691$^{[66]}$ & - & 1 & 1 & 2 & 0.093 \\ 
XTE J1751--305 & 0.71$^{[67]}$ & P & - & 52367$^{[68]}$ & - & 1 & 0 & 1 & $\leq$0.047 \\ 
\emph{Swift} J1756.9--2508 & 0.91$^{[69]}$ & P & - & 54258$^{[70]}$ & - & 1 & 0 & 1 & $\leq$0.047 \\ 
HETE J1900.1--2455 & 1.39$^{[71]}$ & P & - & 53535$^{[72]}$ & - & 1 & 0 & 1 & $\leq$0.047 \\ 
1A 1744--361 & 1.62$^{[73]}$ & D & - & 42837$^{[74]}$ & >=2$^{[74,75]}$ & 4~(1) & +1$^*$$^{[76]}$ & 5$^{**}$ & 0.234 \\ 
SAX J1808.4--3658 & 2.01$^{[77]}$ & P & - & 50338$^{[78]}$ & - & 5 & 2 & 7 & 0.327 \\ 
GS 1826--238 & 2.09$^{[79,79,80]}$ & M & - & 47412$^{[81]}$ & 1$^{[81]}$ & 0 & 0 & 1 & $\leq$0.035 \\ 
IGR J00291+5934 & 2.46$^{[82]}$ & P & - & 53341$^{[83]}$ & - & 0 & 1 & 1 & $\leq$0.047 \\ 
IGR J17511-3057 & 3.47$^{[84]}$ & P & - & 55086$^{[85]}$ & - & 1 & 0 & 1 & $\leq$0.047 \\ 
EXO 0748--676 & 3.82$^{[86,87]}$ & D+M & G5-9V?$^{[88]}$ & 46111$^{[89]}$ & 1~(0)$^{[89]}$ & 0 & 0 & 1 & $\leq$0.031 \\ 
XTE J2123--058 & 5.96$^{[90,91]}$ & A+M & K5-7V$^{[91,92]}$ & 50991$^{[93]}$ & - & 1 & 0 & 1 & $\leq$0.047 \\ 
MXB 1659--298 & 7.11$^{[94]}$ & D & K2(or~later)V?$^{[95]}$ & 43053$^{[96]}$ & 1$^{[96]}$ & 2 & 0 & 3 & 0.074 \\ 
GRS 1747--312 & 12.36$^{[97]}$ & D & - & 48143$^{[98]}$ & >=1$^{[98]}$ & 12 & 0 & 12$^{**}$ & 0.561 \\ 
4U 1608--52 & 12.89$^{[99]}$ & M & ForG?$^{[99]}$ & 40687$^{[100]}$ & 10~(10)$^{[100,101,102,103,104]}$ & 13~(15) & 5 & 28 & 0.594 \\ 
Cen X-4 & 15.1$^{[105,42]}$ & A+M & K3-7V$^{[106,107,108]}$ & 40411$^{[109]}$ & 2~(2)$^{[109,110,111]}$ & 0 & 0 & 2 & 0.042 \\ 
MAXI J0556--332 & 16.43~or~9.75$^{[112]}$ & A & - & 55572$^{[113]}$ & - & 1 & 0 & 1 & $\leq$0.047 \\ 
Aql X-1 & 18.95$^{[114,115]}$ & M & K7V$^{[116]}$ & $\sim$40412$^{[117]}$ & 19~(13)$^{[117,118]}$ & 19~(11) & 5 & 43 & 0.897 \\ 
4U 0042+32 & 278.4$^{[119]}$ & D$^{***}$ & G?$^{[120]}$ & 40621$^{[121]}$ & 1~(1)$^{[121]}$ & 0 & 0 & 1 & $\leq$0.021 \\ 
GRO J1744--28 & 284.02$^{[122]}$ & P & (GorK)III$^{[123]}$ & 50053$^{[124]}$ & 1$^{[124]}$ & 1~(2) & 1 & 3 & 0.140 \\  
\end{longtable*}

Notice that some outbursts that started before the launch of RXTE operated were also detected by RXTE/ASM  (e.g. the outburst of GRO J1744-28 in late 1995). We took these outbursts as the outbursts before the RXTE era.
For those quasi-persistent sources that showed an outburst lasting more than 10 years, such as GRS 1915+105, GS 1826--238, HETE J1900.1--2455 (\citealt{Degenaar+etal+2017}) and EXO 0748--676 (\citealt{Wolff+etal+2008}), we took them as one outburst from these sources. 

1A 1744--361 and GRS 1747--312 were discovered before the RXTE era. The peak fluxes of their outbursts were slightly higher than 50 mcrab in the soft X-ray band and were extremely weak in the hard X-ray band. In order to estimate the outburst rate of these two sources accurately, we only counted their outbursts during or after the RXTE era and calculated the outburst rates in the corresponding period. Additionally, 1A 1744--361 is not in the monitorin-37 in NGC 6441 (\citealt{atel+5301}). We counted this outburst too. 

The numbers of outbursts in our samples are listed in Table~\ref{table1}. For comparison, we listed the number of outbursts before the RXTE era based on Table 7 of CSL97
and the number of outbursts during the RXTE era from Table 1 of YY15 .  Notice that we only took the outbursts above 50 mcrab from CSL97.  We also listed the number of outbursts for BH LMXBTs based on Table 14 of TSHG16 behind our results. 

There are some differences between our results and those in YY15 because we also used \emph{Swift}/BAT data to search for the outbursts and applied a threshold of 50 mcrab instead of 100 mcrab. 
Because YY15 had included several secondary outbursts of 4U 1608--522,  their number of outbursts for 4U 1608--522 is more than our result. Furthermore, we excluded two outbursts from GRO J0422+32 in 1993 that are shown in TSHG16, because these two outbursts could only be detected in the optical band rather than the X-ray band (\citealt{Callanan+etal+1995,Shrader+etal+1997}). The flux of the long-duration outburst in EXO 0748--676 was just 38 mcrab from the result of CSL97 but the flux of this outburst increased to above 50 mcrab in the RXTE era, so we included this outburst in our samples.

\subsection{The outburst rates}
The outburst rate was calculated by $R_{\rm out}=N_{\rm out}/T$, where $N_{\rm out}$ is the number of outbursts since the LMXBT was discovered with X-ray observations and $T$ is the time interval in years since its discovery until MJD 57900. 
For those LMXBTs discovered during the operations of RXTE/ASM, \emph{Swift}/BAT or MAXI, $T$ is set to the interval from the date when RXTE/ASM started operation until MJD 57900, due to similar sensitivities of these missions. In general, for those LMXBTs in which only one outburst has been detected above the threshold during the entire history of X-ray observations, we can only set the upper limit on the outburst rate. 

Notice that some outbursts from these LMXBTs may have been missed owing to poor sensitivity and coverage in early X-ray missions.

\begin{figure*}
  \centering
	\includegraphics[width=18cm]{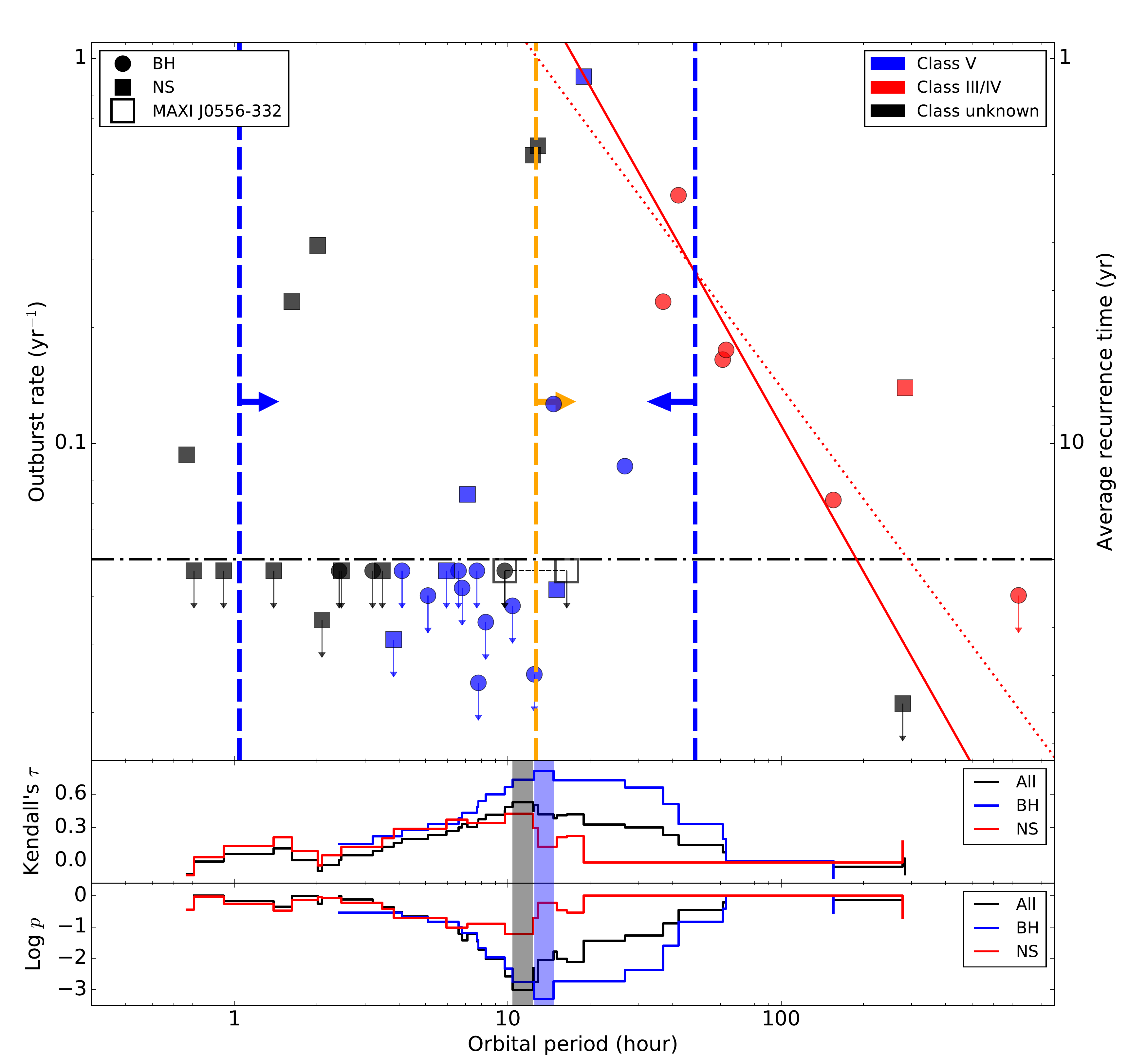}
    \caption{\textbf{Top panel:} The outburst rate as a function of orbital period for BH (circles) and NS (squares) LMXBTs.
    Those LMXBTs from which only one outburst has been detected in historical X-ray observations are labeled with downward-pointing arrows.
NS X-ray binary MAXI J0556--332, of which the orbital period is uncertain, is marked as red open squares at 9.75 and 16.43 hr.
Colors represent the luminosity class of the donor: V (blue), III/IV (red), and unknown (black).
The horizontal dot-dashed line indicates the outburst rate of 0.05~yr$^{-1}$.
The vertical orange dashed line indicates the derived lower limit of the orbital period for a BH X-ray binary to harbor a donor of a subgiant, as discussed in Section~\ref{critical period}.
The vertical blue dashed lines indicate the derived lower limit of orbital period for a BH X-ray binary to harbor a 0.1 ${\rm M_{\odot}}$ main-sequence donor star and the upper limit to harbor a 3 ${\rm M_{\odot}}$ main-sequence star, respectively. 
The best-fitting power-law models to the BH subsamples with a donor of luminosity class III/IV are represented by the red solid/dotted lines (excluding/including GRS 1915+105).
\textbf{Middle panel and bottom panel:} The Kendall's $\tau$ and corresponding $p$-value for various boundaries of the orbital period. The black, blue, and red solid lines represent the results from all, BH and NS samples, respectively. The gray and blue areas represent the maximum $\tau$ and minimum $p$ for all and BH samples, respectively.
}

    \label{fig1} 
\end{figure*}

\section{Results}\label{result}

We plotted the outburst rate as a function of the orbital period for BH and NS LMXBTs in Figure~\ref{fig1}. 
The outburst rates of those sources with good measurements are in the range from less than 0.03 yr$^{-1}$ to near once a year. It is surprising that the outburst rate vs. orbital period relation shows a systematic difference in the systems with the orbital periods below and above $\sim$12 hr. An obvious feature is that almost all samples with the orbital periods below $\sim$12 hr showed only one outburst during the X-ray observational history. Those transients with frequent outbursts, e.g. Aql X-1 and 4U 1608-52, tend to have an orbital period above 12 hr. There is a very large dispersion (or a jump) in the relation between the outburst rate and the orbital period at around $\sim$12 hr. Furthermore, for the samples with a luminosity class of III/IV, the outburst rates show a broad trend to decrease with the orbital period.

\subsection{Different outburst rates in the LMXBTs below and above $\sim$12 hr}
\label{result_difference}
The outburst rate could be associated with the orbital period. The difference in the outburst rate below and above $\sim$12 hr is very likely to originate from a transition of an ordinal variable (e.g. the level of orbital period) or a potential categorical variable (e.g. evolutionary stage). Hence, we divided the all samples into two subsamples, i.e. long-period subsamples and short-period subsamples, based on the selection of a boundary of the orbital period. Since the outburst rates of some samples were obtained as upper limits, we applied Kendall's $\tau$ correlation coefficient from survival analysis to investigate the association between the outburst rate and the level of the orbital period.

As shown in Figure~\ref{fig1}, we have calculated the $\tau$ and the $p$-value for all possible boundaries by using the function \emph{cenken} in CRAN package NADA within the R statistical software (\citealt{Nondetects+2005,Akritas+1995}). We have employed the tie correction to keep $\tau$ in the range [-1, 1] when a large number of tied pairs arise in our data.
For overall samples, the peak ($\tau=0.53$) appears at 10.4--12.4 hr with $p=1.0\times10^{-3}$ ($\sim3.3\sigma$), which means that the outburst rate is associated with the level of the orbital period. On the other hand, we also calculated Kendall's $\tau$ for BH and NS subsamples individually. The Kendall's $\tau$ peaks at 12.5--14.7 hr ($\tau=0.81$) with $p=5.1\times10^{-4}$ ($\sim3.5\sigma$) for BH subsamples and 9.8--12.4 hr ($\tau=0.43$) with $p=0.061$ ($\sim1.9\sigma$) for NS subsamples.
Hence, the outburst rate and the orbital period are strongly dependent on each other in the BH systems. The association is even stronger than that in the overall samples. 

In addition, we applied Fisher's exact test to examine again the significance of the association between the level of the outburst rate (frequent or infrequent) and the level of orbital period (long or short). The boundaries of the orbital period were set to the values that take the Kendall's $\tau$ to peak values above. 
In order to put all upper limits into the class of ``infrequent outburst'', the boundary of the outburst rate should be set to a value larger than 0.047 yr$^{-1}$ .
And the boundary should divide the samples into two subsamples that are as equal as possible. Hence, we set the boundary to 0.05 yr$^{-1}$.
Based on the selection of these boundaries, the Fisher's exact test gives $p=8.67\times10^{-4}$ ($\sim3.3\sigma$) for overall samples, $p=1.0\times10^{-4}$ ($\sim3.9\sigma$) for BH subsamples, and $p=0.18$ (insignificant) for NS subsamples. These statistical significances are roughly consistent with those obtained by applying Kendall's $\tau$ correlation coefficient above.

\subsubsection{Possible selection effect}
Theory predicts a positive correlation between the orbital period and the outburst peak luminosity in LMXBTs (\citealt{King+Ritter+1998,Shahbaz+etal+1998}), which is also confirmed by the X-ray observations (\citealt{Wu+etal+2010}). The LMXBTs with a shorter orbital period tend to have a lower outburst peak luminosity, so quite a number outbursts from the short-period subsamples may be missed owing to the flux limit (50 mcrab) that we set to form our samples.  We need to address a potential selection effect and further investigate whether the systematic difference between long-period and short-period LMXBTs is affected by the flux criteria. 

The flux criterion, 50 mcrab, corresponds to about $2\times10^{37}$ erg s$^{-1}$ in 3--200 keV energy band for sources with a typical distance of 8 kpc by assuming that their X-ray spectra are roughly similar to those of the Crab. 
\cite{Wu+etal+2010} have fitted the relation between the orbital period and the peak luminosity in a large sample with a straight line and got $\log{L_{\rm peak}/L_{\rm Edd}}=(-1.80\pm0.11)+(0.64\pm0.08)\log{P_\mathrm{orb}}$. 
We can infer that the outbursts in a BH system (if $M_{\rm BH}=10~{\rm M_\odot}$) with an orbital period less than $\sim$0.9 hour and the outbursts in a NS system (if $M_{\rm NS}=1.4~{\rm M_\odot}$) with an orbital period less than $\sim$20 hr could be missed by our selection.
Because the orbital periods of BH subsamples, listed in Table~\ref{table1}, are all longer than 2 hr, the outburst rates measured in BH systems should not be affected by the flux criterion applied. 
Therefore, the association between outburst rate and orbital period in the BH subsamples is reliable.

On the other hand, the orbital periods of most NS subsamples, as listed in Table~\ref{table1}, are shorter than 20 hr, which means that the selection effect in NS LMXBTs cannot be ruled out. The 20 hour limit would correspond to 0.9 hr if the source distance is changed to 3 kpc, so some nearby sources (e.g. Cen X-4) will not be affected. In short, we cannot conclude that there is a relation between the outburst rate and the orbital period in NS LMXBTs.

\subsection{The correlation between the outburst rate and the orbital period in the systems with a luminosity class III/IV donor star}
\label{result_correlation}
As shown in Figure~\ref{fig1}, for the samples with a luminosity class of III/IV, the outburst rate shows a broad trend to decrease with the orbital period. 

We applied a Spearman's rank correlation coefficient
to test the correlation between the outburst rate and the orbital period in
the subsamples with a luminosity class of III/IV, and yielded a coefficient of $-0.89$ at a significance of 99.4\%(2.8$\sigma$), which indicates that the outburst rate and the orbital period are anticorrelated in these LMXBTs. We were going to further investigate the correlations for the NS and BH, respectively.
However, since only one NS sample was identified as harboring a luminosity class III/IV donor star, we then tested the correlation in the black hole subsamples with a luminosity class III/IV donor star and yielded the same coefficient of $-0.89$ at a slightly lower significance of 98.2\%~(2.4$\sigma$). 
And we are not sure that this negative correlation is applicable in the LMXBTs with orbital period shorter than 30~hr, since our subsamples with a luminosity class III/IV donor star all have an orbital period longer than 30~hr.
We will explore the possible explanation of this negative correlation in Section~\ref{discussion_correlation}.

\section{Discussion}

As introduced in Section~\ref{introduction}, the mass transfer rates are very different between case A and case B mass transfers. In Section~\ref{critical period}, we will discuss the critical orbital period for case B mass transfer based on the evolutionary tracks of single stars. The critical orbital period is consistent with the boundary orbital period, which separates the samples into very different levels of outburst rates. Hence, the systematic difference of outburst rates below and above $\sim$12 hr can be associated with different evolutionary stages of the donors.
From popular accretion theories, e.g. the DIM, the mass transfer rate affects the quiescence time ($\approx$ recurrence time). We discuss the relation between the quiescence time and the mass transfer rate first and then discuss the orbital period range corresponding to the case A and case B mass transfers in the following subsection.

\subsection{About the quiescence time}
\label{discussion_quiet}
In theoretical models (e.g. the standard DIM), quiescence time is usually mentioned rather than the recurrence time of outbursts. The recurrence time ${t}_{\rm rec}$ can be expressed as the sum of the quiescence time and the duration of an outburst, namely, $t_{\rm rec}=t_{\rm qui}+t_{\rm out}=t_{\rm qui}/(1-\tau)$, where $\tau$ represents the duty cycle. TSHG16 shows that the average duty cycle of transient BHBs is only 7.3\%. This means that approximately the recurrence time can be taken as the quiescence time in BH systems.

As mentioned in the introduction section, the quiescence time can be estimated by the shorter one of the mass-accumulation time and the viscous drift time in the standard DIM (\citealt{Osaki+1995,Osaki+1996}). However, the standard DIM is challenged by the long quiescence time observed, which can exceed 20--30 years in some short-period LMXBTs (typically $P_{\rm orb} < 12$ hr), because the viscous drift time is shorter than one year in these systems if the viscosity parameter $\alpha$ is given by the ``standard value'' (see \citealt{Menou+etal+2000}). \cite{Hameury+etal+1997} suggested that the disk would be truncated at a sufficiently large radius, which will increase the quiescence time, because the viscous drift time is only valid for $r_{\rm out}/r_{\rm in}\gg 1$ (\citealt{Lasota+2001}). In theory, the accretion disk can be truncated by the evaporation mechanism \citep{Meyer+Meyer-Hofmeister+1994} or the magnetic field \citep{Livio+Pringle+1992}. Therefore, the quiescence time is usually estimated as the mass-accumulation time in LMXBTs (e.g. \citealt{Menou+etal+1999+faint,Knevitt+etal+2014}). 

According to \cite{Lasota+2001}, the quiescent time can be expressed as 
\begin{equation}  
t_{\rm qui}\approx \frac{\epsilon M_{\rm D,max}}{\dot{M}_{\rm T}-\dot{M}_{\rm in}},
\label{equation_quiec}
\end{equation}
where $M_{\rm D,max}$ is the maximum disk mass in the quiescence, $\epsilon$ represents the fraction of mass loss of the accretion disk during outburst, $\dot{M}_{\rm T}$ is the mass transfer rate from the donor, and $\dot{M}_{\rm in}$ is the accretion rate at the inner edge of the disk. In the standard DIM, $\dot{M}_{\rm in}\ll\dot{M}_{\rm T}$ holds during the quiescence.

The maximum disk mass can be estimated by assuming that the accretion disk before an outburst is filled up to the critical surface density. The integration of the critical surface density to the disk area yields $M_{\rm D,max}\propto r_{\rm out}^{3.15}M_1^{-0.38}$ \citep{Hameury+etal+1998,Menou+etal+1999+faint,Lasota+2001,Knevitt+etal+2014}. On the other hand, \cite{Paczynski+1977} have derived the maximum size of an accretion disk in a close binary system by assuming vanishingly small pressure in the disk. As a result, the maximum size of the disk is a function of mass ratio and orbital period of the binary system. According to \cite{Lasota+etal+2008}, the outer disk radius is positively related to the orbital period and the mass of the compact object, namely, $r_{\rm out} \propto P_{\rm orb}^{2/3}M_1^{1/3}$. Hence, we obtained the relation between the maximum disk mass and the orbital period, i.e. $M_{\rm D,max} \propto P_{\rm orb}^{2.1}M_1^{2/3}$. Since the maximum disk mass increases with the orbital period at a given $M_1$, lower recurrence time in long-period LMXBTs ($P_{\rm orb} \gtrsim 12$ hr) cannot be caused by the term $M_{\rm D,max}$ unless the viscosity parameter $\alpha$ is much lower than the ``standard value'' in the short-period LMXBTs (\citealt{Menou+etal+2000}).

Instead of adjusting the parameters (e.g. $\alpha$ and $\epsilon$) of the theoretical models, a more intuitive explanation of our results is that the mass transfer rate $\dot{M}_{\rm T}$ is systematically different between long-period and short-period LMXBTs owing to different evolutionary stages.

\subsection{The critical orbital period for case B mass transfer}
\label{critical period}
\begin{figure}
  \centering
	\includegraphics[width=8.5cm]{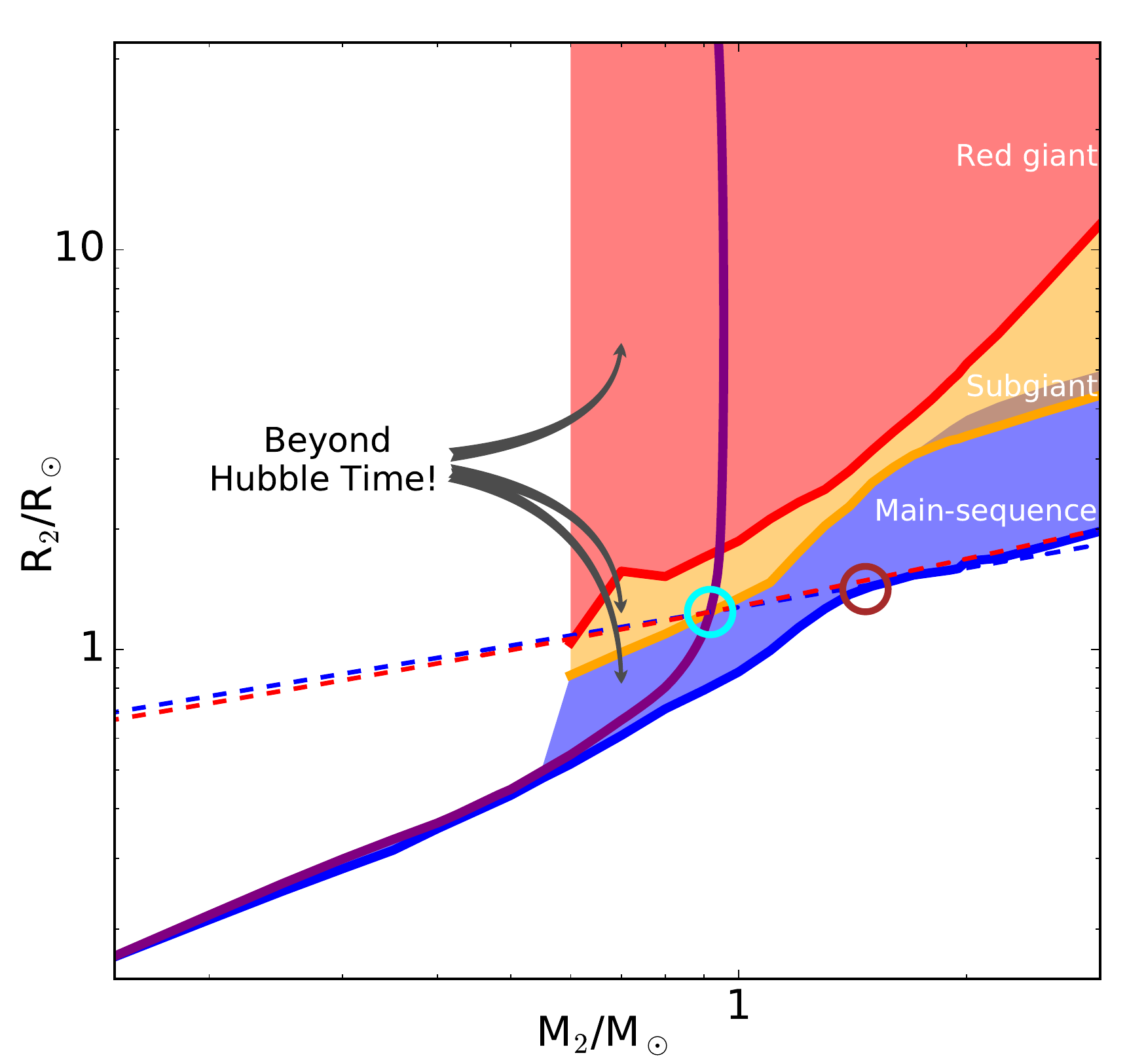}
    \caption{
   Mass-radius relations for main-sequence stars (blue area), subgiants (orange area) and giants (red area) in solar metallicity (Z=0.019). The blue, orange, and red solid lines represent the mass-radius relations of the ZAMS, the TAMS and the beginning of the red giant branch, respectively. The purple solid line indicates the mass-radius relation in a 14.1 billion years ($\approx$ Hubble time) isochrone. The area to the left of this line exceeds the Hubble time. We used a cyan circle to highlight the intersection of the isochrone and the TAMS and also a brown circle to highlight the inflection of the ZAMS. The dashed lines represent the mass-radius relations of a Roche lobe filling donor star to the critical orbital period if $M_1=10~{\rm M_{\odot}}$ (blue) or $M_1=1.4~{\rm M_{\odot}}$ (red).
   }
    \label{fig2} 
\end{figure}

As introduced in Section~\ref{introduction}, case B mass transfer can be performed only if the LMXBTs have an orbital period greater than a critical orbital period $P_{\rm orb,crit}$. We can infer the critical orbital period based on the evolutionary tracks of single stars and the assumption that the radius of a single star just equals to the Roche lobe radius. As shown in Figure~\ref{fig2}, blue, orange, and red areas represent the mass-radius relation (MRR) of main-sequence stars, subgiants, and red giants in solar metallicity (Z=0.019), respectively. These evolutionary tracks are taken from the Podova Stellar Evolution Database \citep{Girardi+etal+2000}.  
The purple solid line is the MRR in a 14.1 billion yr ($\approx$ Hubble time) solar-metallicity isochrone taken from \cite{Girardi+etal+2000}. Obviously, the age of a star cannot exceed Hubble time. Hence, we can obtain the smallest radius and the lowest mass of a subgiant from the intersection of the isochrone and terminal-age main sequence (TAMS, the orange solid line). Obviously, the LMXBT system with the critical orbital period must just harbor such a subgiant. For the solar metallicity Z=0.019, the smallest radius is 1.2~$\rm R_{\odot}$ and the lowest mass is 0.9~$\rm M_{\odot}$. On the other hand, the MRR of a Roche lobe filling donor star can be determined when the orbital period and the mass of the compact object are both known. For an X-ray binary for which the orbital period equals the critical orbital period, its MRR must pass through the intersection of the isochrone and the TAMS. Therefore, we obtain $P_{\rm orb,crit}=$12.9 hr for $M_1=10~{\rm M_{\odot}}$ (blue dashed line) and $P_{\rm orb,crit}=$12.6 hr for $M_1=1.4~{\rm M_{\odot}}$ (red dashed line). Obviously, the critical orbital period is insensitive to the mass of the compact object. Notice that we did not take into account the cases of mass loss by stellar winds or mass transfer. These effects can lead to lower mass of the (sub)giants.

We also took into account the cases under a wider range of metallicity, from Z=0.0004 to 0.03. We obtained that the lower limits of the critical orbital period are 12.7 hr for typical BH systems ($M_1=10~{\rm M_{\odot}}$) and 12.4 hr for typical NS systems ($M_1=1.4~{\rm M_{\odot}}$). This means that \emph{we cannot find any LMXBT system with the orbital period below 12.4 hr in which the donor performs case B mass transfer}. 
Therefore, the mass transfer rates in the systems with periods below 12.4 hr are very low when they all only perform case A mass transfer, unless their orbital periods are short enough for significant angular momentum loss by gravitational wave radiation.
From Eq.~(\ref{equation_quiec}), a lower mass transfer rate will lead to a longer quiescence time (or recurrence time). Although $M_{\rm D,max}$ is also dependent on the orbital period, $M_{\rm D,max} \propto P_{\rm orb}^{2.1}$ cannot offset the gap of several hundred (or even thousand) times between case A and case B mass transfer rates unless the orbital period is much shorter than those case B binaries (typically $P_{\rm orb} < 2 $ hr). Therefore, most LMXBTs below $\sim$12 hr show a very long recurrence time, which can be very different from those LMXBTs above $\sim$12 hr.

Back to our results, for the BH systems, $P_{\rm orb,crit}=12.7$ hr is consistent with the boundary of orbital period 12.6--14.5 hr that takes the Kendall's $\tau$ to a maximum value in Section~\ref{result_difference}. All BH subsamples below 12.7 hr show a very low outburst rate (<$0.05~{\rm yr}^{-1}$) and almost all BH subsamples above 12.7 hr show a relatively higher outburst rate ( >$0.06~{\rm yr}^{-1}$ at least), except for a sample (GRS 1915+105) with a very large orbital period. 

On the other hand, the critical orbital period should be applicable to the NS LMXBTs in theory. However, we did not find a significant difference in outburst rates for NS subsamples. In fact, the behaviors of some NS LMXBTs are not well understood. For example, the orbital period of GRS 1747-312 is slightly lower than the critical orbital period, which means that the donor of GRS 1747-312 is more likely to be a main-sequence star, but its outburst is very frequent.

\subsection{The wide period range for case A mass transfer}

When the orbital period is above $P_{\rm orb,crit}$, the system can harbor a main-sequence star and a (sub)giant. 
From our samples, the maximum mass of the known main-sequence donors may be $\approx2.5-3.0~{\rm M_{\odot}}$ (\citealt{Orosz+etal+1998,Orosz+2003,Tetarenko+etal+2016}). Therefore, we assumed that the maximum mass of main-sequence donors in LMXBTs is around $3~{\rm M_{\odot}}$.
To harbor a $3~{\rm M_{\odot}}$ main-sequence star with Z=0.0004--0.03, the upper limit of the orbital period can reach 48.4 hr for $M_1=10~{\rm M_{\odot}}$ or 41.6 hr for $M_1=1.4~{\rm M_{\odot}}$. 
The candidate mass range of the main-sequence donor becomes wider when $P_{\rm orb}$ is greater than the period $P_{\rm orb,infle}$ ($\approx P_{\rm orb,crit}$ when Z=0.019) corresponding to the inflection of the MRR of the zero-age main sequence (ZAMS, blue solid line in Figure~\ref{fig2}), because the slope of the ZAMS becomes lower when its mass exceeds around 1.47 ${\rm M_{\odot}}$, which could be caused by the transition of nuclear energy from the proton--proton cycle to the carbon--nitrogen cycle \citep{Demircan+Kahraman+1991}. The $P_{\rm orb,infle}$ can reduce to $\sim$9 hr for those metal-poor donors.
Since $t_{\rm nuc}\propto M_2^{-2.5}$, a massive main-sequence donor (e.g. $2~{\rm M_{\odot}}$) can also contribute slightly substantial (case A) mass transfer (about several times $10^{-9}~{\rm M_{\odot}/yr}$).

Similarly, for an X-ray binary to harbor a very low mass main-sequence star (e.g. $0.1~{\rm M_{\odot}}$), its orbital period must be longer than 1.0 hr if $M_1=10~{\rm M_{\odot}}$ or 1.1 hr if $M_1=1.4~{\rm M_{\odot}}$, which are derived from the MRR of the ZAMS \citep{Tout+etal+1996}. We have also plotted these limits on the orbital period for BHs in Figure~\ref{fig1} in order to intuitively understand the association among outburst rate (or recurrence time), orbital period, and evolutionary stage. From $\sim$12 to $\sim$48 hr, there are both case A binaries and case B binaries. This means that the mass transfer rates (and also the outburst rate) in this period range will show a very large dispersion with orbital period, which can be seen in our results as shown in Figure~\ref{fig1}. The period range, 12--48 hr, is numerically consistent with the ``bifurcation'' orbital period \citep{Pylyser+Savonije+1988,king+etal+1996,Menou+etal+1999+quiet}, which separates the converging binary systems ($P_{\rm orb}$ decreases with time) and the diverging binary systems ($P_{\rm orb}$ increases with time). 

Notice that, besides the main-sequence and (sub)giant donor, LMXBTs also can harbor a donor of a brown dwarf or a WD. For example, derived from the evolutionary models of \cite{Chabrier+etal+2000}, the upper limit of the period for brown dwarfs is around 4 hr.

\subsection{The outburst rate decreases with the orbital period in the system with a donor of a (sub)giant}
\label{discussion_correlation}
As introduced in Section~\ref{result_correlation}, we have found a negative correlation between the outburst rate and the orbital period in the subsamples with a luminosity class of III/IV donor star. We then discuss this correlation under the framework of the DIM. From Eq.~(\ref{equation_quiec}), since the $\dot{M}_{in}$ is negligible in quiescence, the outburst rate (or recurrence time) is mainly determined by disk mass $M_{\rm D,max}$ and mass transfer rate $\dot{M}_{\rm T}$.

The luminosity classes III/IV and long orbital periods ($P_{\rm orb}\geq 30$ hr) of these subsamples indicate that their donor stars are very likely to be subgiant or giant stars. \cite{King+1988} has derived a relation between the mass transfer rate and the orbital period for those systems with a donor evolved off main sequence, $\dot{M}_{\rm T} \propto P_{\rm orb}^{0.93}M_{2}^{1.47}$. 
And we have obtained $M_{\rm D,max} \propto P_{\rm orb}^{2.1}M_1^{2/3}$ based on the DIM in Section~\ref{discussion_quiet}.
Then, we obtain that $t_{\rm rec}\approx t_{\rm quiec} \propto \epsilon M_{1}^{2/3}M_{2}^{-1.47}P_{\rm orb}^{1.17}$. Hence, in a LMXBT with a donor of a (sub)giant, the recurrence time increases with the orbital period. If the $\epsilon$ is constant, at a given $M_{1}$ and $M_{2}$, the outburst rate decreases with orbital period as a power-law form $R_{\rm out} \propto P_{\rm orb}^{-1.17}$.

Because the theoretical outburst rate (or recurrence time) is dependent on the mass of the compact object and there is only one NS sample in the correlation, we therefore used a power-law model to fit the correlation in the BH subsamples, except GRS 1915+105, the outburst rate of which is only an upper limit. By applying the Cash statistic (\citealt{Cash+1979}), we obtained the relation between the outburst rate and the orbital period as $R_{\rm out}=1.61^{+1.80}_{-0.81}\times(P_{\mathrm{orb}}/\rm 12~h)^{-1.26^{+0.47}_{-0.53}}~{\rm yr^{-1}}$ (see Figure~\ref{fig1}).
The uncertainties correspond to $\Delta C=1$ in the Cash statistic.
For a comparison, we also fitted the relation; GRS 1915+105 was included, and its outburst rate was taken as a measured value.
Then, we obtained $R_{\rm out}=1.07^{+0.80}_{-0.42}\times(P_{\mathrm{orb}}/\rm 12~h)^{-0.96^{+0.32}_{-0.37}}~{\rm yr^{-1}}$,
which has a slightly flatter slope.

The best-fitting power-law index is roughly consistent with the above estimation from the DIM, no matter whether GRS 1915+105 is included or not. In conclusion, the negative correlation in these BH subsamples supports the idea that the DIM plus the Case B mass transfer would cause a longer recurrence time in the system with a longer orbital period. 
However, the spread of outburst rates around the best-fitting relation is broad. 
The large scatter of this negative correlation indicates that other parameters (such as $M_{1}$ and $M_{2}$) besides the orbital period affect the outburst rate (or recurrence time) as well.

In our samples, two other BH LMXBTs (GRS 1716--249 and 4U 1543--47) are worth to be mentioning. Their donors are luminosity classes V, but their orbital periods are in the mixture region for a BH LMXBT with a main sequence or a (sub)giant donor star (see the orange and blue dashed vertical lines in Figure~\ref{fig1}).
We calculated the Spearman correlation coefficient as $-0.33$ with $p=0.38$ (insignificant), in the samples consisting of GRS 1716--249, 4U 1543--47 and all subsamples of which the luminosity classes are III/IV. The test indicates that these two sources do not follow the negative correlation that we found in luminosity class III/IV subsamples. The possible explanation is that these two sources may not harbor (sub)giant donors, or they do not follow the mass transfer rates derived by \cite{King+1988}.

On the other hand, the essential mechanism to trigger outbursts is the same between NS LMXBTs and BH LMXBTs in the DIM. The correlation also should be available for the NS LMXBTs of which the donor is a (sub)giant. In Figure~\ref{fig1}, some NS samples (e.g. Aql X-1, 4U 1608--52 and GRS 1747--312) are in line with the best-fitting model for the BH subsamples. 
Therefore, we suggest that these systems could all harbor a subgiant donor. Notice that although the spectral type of the donor in Aql X-1 is identified as K7V (\citealt{Chevalier+etal+1999}), but this will lead to an estimate of the distance of 2.5 kpc, which is much closer than the distance estimated by type I X-ray bursts (\citealt{Galloway+etal+2008,Rutledge+etal+2001}), and it is also difficult to explain how a K7V donor fills a Roche lobe of a 19-hour orbit.


\subsection{The orbital period distribution of low-mass X-ray binaries}
The relation between the recurrence time and the orbital period plays an important role in regulating the orbital period distribution of low-mass X-ray binaries, which is essential to restrict the evolution models of X-ray binaries.

Since the detection probabilities of LMXBTs are strongly dependent on the orbital periods, the observed distribution of LMXBs suffers significant selection effects. The detection probabilities can be estimated based on the peak luminosity, the outburst duration, and the recurrence time in the relation with the orbital period (\citealt{Knevitt+etal+2014,Arur+Maccarone+2018}). As the key factor to build the orbital period distribution of LMXBs, however, the relation between the recurrence time and the orbital period is usually not well determined. For example, from \cite{Knevitt+etal+2014}, they derived that the recurrence time increases continuously with the orbital period when the orbital period is above 2 hr, which is inconsistent with the observed results. Therefore, we suggest that the period distribution of LMXBs should be estimated based on the actual relation.



\section{Summary}
We have investigated the outburst rate in relation to the orbital period for the current LMXBT samples of which the orbital period has been measured.
By applying the Kendall's $\tau$ and the Fisher's exact test, we found that the outburst rates are systematically different between long-period ($P_{\rm orb} \gtrsim$ 12 hr) and short-period ($P_{\rm orb} \lesssim$ 12 hr) subsamples. By analyzing BH and NS subsamples separately, we found that the systematic difference is still significant for the BH subsamples but not very significant for NS subsamples. We take into account the selection effect due to the positive correlation between the orbital period and outburst peak luminosity in LMXBTs. Then, we rule out the selection effect on BH systems, but we cannot rule out the effect on NS systems.

We infer that the mass transfer rates are responsible for the systematic difference of outburst rates since the DIM suggested that the mass transfer rate is a key factor affecting the quiescence time of LMXBTs and the mass transfer rates could be very different in long-period and short-period LMXBTs owing to different evolutionary stages of donors. From the evolutionary tracks of stars, we derived that the critical orbital period to harbor a (sub)giant donor or perform case B mass transfer is 12.7 hr for BH LMXBTs and 12.4 hr for NS LMXBTs. 
 The critical orbital period is consistent with the boundary of the orbital period, which separates the samples into very different levels of outburst rates.
This can explain why there is a systematic difference of outburst rates between the systems with orbital periods below and above $\sim$12 hr. 

We also determined a wide orbital period range, 1.1 to $\sim$48 hr, for X-ray binaries to harbor a main-sequence donor or perform case A mass transfer. Since this period range overlaps the period range for X-ray binaries to harbor a (sub)giant donor, we expect that the outburst rate of LMXBTs in the orbital period range from $\sim$12 to $\sim$48 hr will show a very large dispersion, which can be verified in our results.

Furthermore, we found a negative correlation between the outburst rate and the orbital period in the subsamples for which the luminosity class is III/IV. The high luminosity classes and long orbital periods of these samples mean that their donor stars are very likely to be subgiants or giants. For those LMXBTs with a donor star evolved off the main sequence, the relation between the outburst rate (or recurrence time) and the orbital period follows a power-law form under the framework of the DIM, and the power-law index is roughly consistent with our best-fitting results for the BH subsamples.

On the other hand, the essential mechanism that triggers outbursts is the same between NS and BH LMXBTs in the current theory of the DIM. We cannot investigate the correlation for NS LMXBTs, since there is only one NS LMXBT (GRO J1744--28) for which the luminosity class is III/IV.
And some NS LMXBTs (e.g. 4U 1608--52 and GRS 1747--312 ) seem in line with the negative correlation, although the luminosity classes of their donors are not credibly identified yet. We suggest that a similar negative correlation should exist in these NS LMXBTs harboring a (sub)giant donor.

\acknowledgments
We would like to thank RXTE and \emph{Swift} Guest Observer Facilities at NASA Goddard Space Flight Center for providing RXTE/PCA and \emph{Swift}/XRT products and \emph{Swift}/BAT and RXTE/ASM transient monitoring results. This research has made use of MAXI data provided by RIKEN, JAXA, and the MAXI team. W.Y. would like to acknowledge the support by the National Program on Key Research and Development Project (Grant no. 2016YFA0400804), by the National Natural Science Foundation of China under grant No. U1838203 and 11333005 and by the FAST fellowship, which is supported by Special Funding for Advanced Users, budgeted and administrated by Center for Astronomical Mega-Science, Chinese Academy of Sciences (CAMS). Z.H. would like to acknowledge the support by National Natural Science Foundation of China under grant no. 11521303 and no. 11733008. Z.Y. would like to acknowledge the support by the National Natural Science Foundation of China under grant no. 11773055.



\end{document}